\documentstyle[12pt]{article}
\newcommand{\half}{\frac{1}{2}}
\newcommand{\ov}{\overline}

\begin{document}%

\title{\Large \bf
On the concept of the tunneling time}
\author{
{\normalsize\sc M. S. Marinov and Bilha Segev}\\
{\normalsize\it
Department of Physics, Technion-Israel Institute of Technology}\\
{\normalsize\it Haifa 32000, Israel}\\}

\date{March 1996 }

\maketitle
{\begin{abstract}
Asymptotic time evolution of a wave packet describing a 
non-relativistic particle incident on a potential barrier is 
considered, using the Wigner phase-space distribution. The distortion of 
the trasmitted wave packet is determined by two time-like parameters, 
given by the energy derivative of the complex transmission amplitude.
The result is consistent with various definitions of the tunneling time 
(e.g. the B\"{u}ttiker-Landauer time, the complex time and Wigner's
phase time). The speed-up effect and the negative dispersion are 
discussed, and new experimental implications are considered. 
\end{abstract}


\newpage
Quantum tunneling, where a particle has a chance to pass through
classically forbidden regions, was one of the first important applications 
of wave mechanics. The total barrier penetration probability may be 
calculated directly from the stationary Schr\"{o}dinger equation, yet
the process time dependence is a more delicate phenomenon.
The source of the problem is the uncertainty relation: if the incident
momentum was known exactly, the coordinate would be absolutely uncertain,
and one could hardly ask a question about the particle transport.
In principle, the solution is evident\cite{goldberger}: a wave packet
must be prepared beyond the potential domain, far from it,
so that one can afford a wide spread, compatible with a relatively small 
momentum uncertainty. After that, one has to wait long enough until the 
wave packet would penetrate through the barrier and leave it completely,
being splitted in two, forward and backward. However, the problem needs an 
intricate analysis based upon the time-dependent formalism.

The question: ``How much time does the tunneling process take?",  
has been standing since 1932 (e.g. Ref.\cite{maccoll}). Various definitions, 
approaches, experiments and reviews were worked out 
(a partial list is given in Refs.\cite{hartman}-\cite{steinberg}) in 
order to answer that question which is substantial for physical 
applications and for a proper understanding of quantum theory.
 
There are three cardinal approaches to the concept of the tunneling time.
\begin{itemize}
\item Path integrals and the semiclassical approximation 
suggest\cite{fertig,wardlaw2,sokol&connor} 
employing a {\em complex} time $\tau^{\rm C}$.
\item A physical {\em clock} may be itroduced\cite{perestime},
at least at the level of a ``Gedankenexperiment". One way to measure the 
time spent under the barrier is to study the particle spin precession in
an external magnetic field\cite{baz,rybachenko}, which leads to a Larmor
time $\tau^{\rm L}$. Another way is to consider a vibrating barrier
\cite{buttiker&landauer,martin&landauer}, getting the B\"{u}ttiker -- 
Landauer time $\tau^{\rm BL}$. 
\item One can try to trace the behaviour of 
the wave packet in interaction with the barrier
\cite{maccoll,hartman,dumont&marchioro II,steinberg&chiao,steinberg}. 
This approach has two frail points
\cite{buttiker&landauer,leavens&aers,landauer&martin}:
i) a dependence on the initial state preparation, ii) spreading of the
wave packet outside the potential region. 
    \end{itemize}
It was found subsequently \cite{fertig,muga&brouard&sala2,landauer&martin} 
that the definitions of $\tau^{\rm BL}$, $\tau^{\rm L}$, and $\tau^{\rm C}$ 
are not inconsistent.
On the other hand, for central potential scattering, which is described 
in terms of partial-wave phase shifts, Wigner\cite{wigner} (see also in
Ref.\cite{goldberger}, Ch. 8) introduced a {\em phase time} $\tau^{\rm W}$
(Eq. (1) below) and noted its relation to the {\em causality principle}. 
Recent experiments with optical analogs of quantum 
tunneling\cite{chiaoetal-exper}-\cite{krausz} indicate
an importance of the phase time for barrier penetration.

The purpose of this work is to investigate general features of
deformation of wave packets in the process of tunneling through potential 
barriers. It is assumed that the initial momentum uncertainty is small,
and we look at the large-time asymptotics. The result is expressed in 
terms of the (complex) transmission amplitude $A(\kappa)$, which
is obtained by the solution of the stationary Schr\"{o}dinger equation
with the energy $\epsilon=\kappa^2/2m$, the particle mass being $m$.
The {\em total} transmission probability is given by $|A(\kappa)|^2$,
and the change of shape of the coordinate probability distribution
is determined by two time-like parameters related to the energy 
derivatives of the transmission amplitude, 
    \begin{equation}
\tau^{\rm W}\equiv d(\arg A)/d\epsilon,\;\;\;
\tau^{\rm A}\equiv d(\ln |A|)/d\epsilon,
    \end{equation}
where $d\epsilon=vd\kappa$, and $v=\kappa/m$. (We set $\hbar=1$ throughout
the paper.) It is noteworthy that these parameters appear in other 
formulae for the tunneling time, namely,
$\tau^{\rm C}=\tau^{\rm W}-i\tau^{\rm A}$ and $\tau^{\rm BL}=|\tau^{\rm C}|$.
The Wigner phase time $\tau^{\rm W}$ has the same form as in the central 
scattering, where $|A|\equiv 1$, in contrast to the problem concerned.

We shall use the Wigner phase-space quasi-distribution, i.e. the Weyl 
symbol of the density matrix (see e.g. a review in Ref.\cite{hilleryetal}). 
There are two reasons for that: i) this formalism enables one to treat both 
pure and mixed initial states, and quite general types of the wave packets;
ii) it is easier to get rid of irrelevant oscillations of the wave functions.

The time evolution of the initial Wigner function $\rho_0(q,p)$ can be 
given by means of the phase-space evolution kernel\cite{marin91}, which 
represents the fundamental solution of the Landau -- von Neumann equation 
for the density matrix. Namely, for any time $t$ the Wigner function is 
       \begin{equation}
\rho_t(q,p) = \int L_t(q,p;q_0,p_0)\rho_0(q_0,p_0)dq_0dp_0.   
    \end{equation}
We shall consider the scattering problem, where the initial state is 
prepared with an average momentum $P_0$ and has a relatively small
momentum dispersion $\Delta p_0\ll |P_0|$, which are defined as usual,
   \begin{eqnarray}
&&P_0=\int p\rho_0(q,p)dqdp,\\
&&(\Delta p_0)^2=\int (p-P_0)^2\rho_0(q,p)dqdp.\nonumber
   \end{eqnarray}
Similar definitions hold for the central coordinate $Q_0$ and the dispersion
$\Delta q_0$. As follows from the uncertainty relation, 
$\Delta q_0\Delta p_0\ge\half$, and the inequality may be saturated for a 
set of pure (coherent) states. It is assumed that the potential barrier
$V(q)\ge 0$ is located near the origin and has a finite range $D$, vanishing
for $|q|>D$. The initial state must be prepared in the free space, which 
means that $|Q_0|-\Delta q> D$. The results of the scattering are 
observed after the wave packets gets out of the potential region, say, when
$t>2|Q_0|m/P_0$.

In the large time asymptotics, the phase-space evolution kernel for the 
barrier penetration was found\cite{prop} to be a sum of two parts, describing
transmission and reflection, 
   \begin{equation}
L_t(q,p;q_0,p_0) \asymp \delta(p-p_0)T(r_+,p_0)
+ \delta(p+p_0)R(r_-,p_0),
    \end{equation}
where $r_\pm=q_0+t p_0/m\mp q$. (If the barrier would be absent,
one has $T=\delta(r_+)$, $R\equiv 0$, and $L_t$ is just the solution of 
the classical Poisson equation for free motion.) The functions $T$ and $R$ 
have been expressed in terms of integrals involving the transition and 
reflection coefficients, $A(\kappa)$ and $B(\kappa)$, respectively. In 
particular, the integral representation for the transmission propagator is
       \begin{equation}
T(r,p)=\int^\infty_{-\infty}d\sigma
e^{-i\sigma r} A(p+\half\sigma)\ov{A(p-\half\sigma)}.
          \end{equation}   
This representation is manifestly causal. Note that $A(\kappa)$ has
the following general properties \cite{analy}: it is analytical in 
the upper half of the complex plane (having poles at Im $\kappa<0$), 
$\ov{A(\kappa)}=A(-\bar{\kappa})$, and
$\lim_{\kappa\rightarrow\infty}A(\kappa)=1$. Thus the integral in (5)
can be considered as a contour integral in the complex $\sigma$-plane. 
If $q>q_0+vt$, i.e. $q$ would be in advance of the free motion coordinate, 
the integration contour can be moved up to infinity, so $T(r,p)=0$ for
$r<0$. Thus no point of the Wigner distribution is transported faster than
it would be in the absence of the potential barrier.

The information transport is somewhat smeared by the fact that the phase 
space distribution is never too local because of the uncertainty relation.
Let us consider the observable consequences of the causality arguments 
presented above. We shall suppose that the final coordinates are measured,
and the detector does not discriminate between different momenta.
The probability of finding the transmitted particle at $q$, for 
asymptotically large $t$, is given by the following integral,
     \begin{equation}
{\cal P}_t(q) 
=\int^\infty_{-\infty}dp\int^\infty_{-\infty}dr\int^\infty_{-\infty}d\sigma
e^{-i\sigma r}A(p+\half\sigma)A(-p+\half\sigma)\rho_0(q-vt+r,p),
      \end{equation}   
where $v=p/m$. Dealing with this representation, one can make use of
the specific features of $\rho_0$, which is sharp in $p$ and broad in $q$.
Therefore it is reasonable to expand $A(\pm p+\half\sigma)$ in powers of 
$(p-P_0)$. On the other hand, it was found\cite{prop} that $T(r,p)$ 
is an exponentially decreasing function of $r$, so the fact that the 
$r$-dependence of $\rho_0$ is slow may be taken into account. 
A straightforward calculation shows that one can expand the integral in 
powers of ${\Delta p_0}/{P_0}$. To the first order, the result is 
      \begin{equation}
{\cal P}_t(q)\approx |A(P_0)|^2 \left[ {\cal P}^0_t(q) 
+ v_0\tau^{\rm W}\partial{\cal P}^0_t/\partial q
+ v_0\tau^{\rm A}2M_t(q) \right].
      \end{equation}
Here
      \begin{equation}
{\cal P}^0(q)\equiv\int dp\rho_0(q-vt,p),\;\;\;
M_t(q)\equiv\int dp(p-P_0)\rho_0(q-vt,p),
       \end{equation}   
${\cal P}^0_t$ represents free motion of
the initial wave packet including the
usual spreading and $M_t$ is the first moment
of the $p$-distribution, which also appears in the free motion.
The time parameters $\tau^{\rm A}$ and $\tau^{\rm W}$ in (7) should be 
calculated by Eq. (1) at $\epsilon=P_0^2/2m$, and $v_0=P_0/m$. 
Note that the total 
transmission probability is  
    \begin{equation}   
\int dq{\cal P}_t(q)/\int dq{\cal P}^0_t(q)=|A(P_0)|^2,
     \end{equation}
as usual, since the 
two other terms in (7) are eliminated by the integration.
The expansion in powers of $(\Delta p_0/P_0)$ can be performed to all orders 
leading to a sum over higher derivatives and higher 
moments of $\rho_0$ with coefficients which depend on the barrier shape. 

Two terms in Eq. (7) describe a distortion of the coordinate propability
distribution. It is natural to expect that $\tau^{\rm A}>0$, the tunneling
probability is increasing with energy, so the corresponding term is 
responsible for an advance in the distribution maximum.
It is the so-called speed-up effect\cite{dumont&marchioro II}.
One can tell that tunneling filters out low-energy components of the wave 
packet. As to the other term, the sign of the phase time $\tau^{\rm W}$ is 
not definite. (Roughly speaking, it is positive for narrow barriers and 
negative for wide barriers). The corresponding term represents
interference effects, i.e dispersion due to the barrier.
This may result in an additional advance 
(if $\tau^{\rm W}<0$), or in a delay (if $\tau^{\rm W}<0$).

Sometimes the effect of $\tau^{\rm W}$ is dominating. That is the case when
$|A(\kappa)|$ is constant, as for the example of central scattering,
considered by Wigner\cite{wigner}. In recent experiments with photon wave 
packets\cite{chiaoetal-exper}, the transmission probability was almost 
insensitive to the wavelength within the light frequency band, so that 
$\tau^{\rm A}\ll\tau^{\rm W}$. If these parameters are of the 
same order of magnitude their relative influence may be determined by 
the distance of the detector from the barrier. As the distance (i.e. the time
$t$ ) is increasing the speed-up effect prevails over
the phase-time effect, since the slope of the free probability distribution 
is getting down because of the wave-packet spreading.

Let us consider a simple example where the initial phase-space 
distribution is Gaussian,
       \begin{equation}  
\rho_0(q,p) =C\exp\left[-\frac{(p-P_0)^2}{2(\Delta p_0)^2}
-\frac{(q-Q_0)^2}{2(\Delta q_0)^2}\right].
        \end{equation}
Here $C=(2\pi \Delta p_0\Delta q_0)^{-1}$ is the normalization constant, and
$\Delta p_0\Delta q_0=\half$ if the state is pure. The calculations are 
straightforward now, 
     \begin{eqnarray}   
&&{\cal P}_t^0(q) = \sqrt{2\pi}\Delta q C\exp
\left[-\frac{(q-Q)^2}{2(\Delta q)^2}\right] ,\\
&&M_t(q)=t\frac{(q-Q)(\Delta 
p_0)^2}{m(\Delta q)^2}{\cal P}_t^0(q),\nonumber\\ 
&&Q\equiv Q_0+ t v_0, \;\;\;
(\Delta q)^2 \equiv(\Delta q_0)^2  +(t \Delta p_0/m)^2  . \nonumber
      \end{eqnarray}
In the essential domain, the coordinate distribution given by Eq. (7) is 
     \begin{equation}   
{\cal P}_t(q)=|A(P_0)|^2{\cal P}^0_t(q) 
\left[1+v_0\tau_0(q-Q)/(\Delta q)^2\right] ,
      \end{equation}
where $\tau_0=2t\tau^{\rm A}(\Delta p_0)^2/m-\tau^{\rm W}$, which may 
change its sign with time. The maximum of the transmitted distribution 
is shifted in advance of the that for the free propagation by
     \begin{equation}   
\Delta Q=2v_0\tau_0/\left(\sqrt{1+\zeta^2}+1\right)
      \end{equation}
where $\zeta=2 v_0\tau_0/\Delta q$. In the limit of 
$\Delta p_0\rightarrow 0$, $\Delta q\rightarrow\infty$, one has 
$\zeta\ll 1$, and $\Delta Q\approx v_0\tau_0$. Besides, for
$(\Delta p_0)^2/m\ll\tau^{\rm W}/\tau^{\rm A}t$, i.e. if the time is not
too large (the detector is not too far from the barrier) we get
$\tau_0\approx-\tau^{\rm W}$ and $\Delta Q\approx -v_0\tau^{\rm W}$ 
in agreement with Wigner's prediction and recent 
experiments\cite{chiaoetal-exper}. For large positive $\tau_0$ 
(which may happen at large $t$), the shift of the maximum is 
bounded by the width of the final distribution. Besides,
the transmitted distribution is contracted by interaction with the barrier,
an effect which is called negative dispersion.

Equality (7) may be applied to any other distribution, not necessarily 
Gaussian. An example considered in recent microwave simulations of  
tunneling\cite{ranfagni} is a step function. That distribution cannot be 
realized in quantum mechanics, but it may be considered as a test function.
The shift of the half-height point with respect to the ``free 
propagation" has been also calculated from Eq. (7). The result is 
$v_0\tau_h$, where $\tau_h=t\tau^{\rm A}(\Delta p_0)^2/m-\tau^{\rm W}$. 
Here the speed-up effect is half of that in the Gaussian case, Eq. (12).
This is consistent with the causality arguments; the wave-packet front 
does not move faster because of a potential barrier.
The motion of the peak, as well as the motion of the half-height point, 
have been measured experimentally\cite{chiaoetal-exper}-\cite{krausz}.

Qualitatively, nonrelativistic particle movement is similar to propagation
of light signal through a reflecting stack, like in the experiment of the
Berkeley group\cite{chiaoetal-exper}. The interpretation of the results
based upon the light interference picture leads to qualitatively similar 
results\cite{yoni}. In order to make a quantitative prediction, one
has to calculate the complex transmission coefficient $A(\kappa)$.
If the optical barrier may be prepared with a strong frequency dependence,
the effect of the ``amplitude time" $\tau^{\rm A}$  would be observable, 
besides the ``phase time" $\tau^{\rm W}$. Remarkably, the result of the 
measurements would depend on the distance between the detector and the 
barrier region. 

It should be emphasized, in conclusion, that the wave packet shape must 
be taken into the consideration of tunneling through potential barriers. 
Causality is not violated of course, but it manifests itself 
indirectly in terms of deformation of the wave packet. The phase-space
formalism, introduced by Wigner, is quite appropriate for the 
investigation, and the main corrections for a wave packet with a fairly 
definite momentum are given in Eq. (7). The resulting effect is an 
interplay of those involving a couple of time-like parameters, defined by 
Eq. (1) and owing to the momentum dependence of the complex transmission 
amplitude. Thus, the complex time $\tau^{\rm C}$ introduced 
previously\cite{fertig,sokol&connor} appears actually in the final
distribution; its real part 
is coupled to the coordinate derivative of 
the freely propagating distribution 
and its imaginary part is coupled to the first 
moment of the momentum distribution.

{\em Acknowledgements.} We are obliged to C. Bender, N. Brenner, 
Sh. Gurvitz, M. Moshe, M. Revzen, A. Ron, and J. Zak for their interest 
in this work and helpful comments. The research was supported by the Technion 
V. P. R. Fund - Bernstein Research Fund, and G. I. F.


\end{document}